\documentclass[american,aps,pra,reprint, superscriptaddress]{revtex4-1}
\usepackage{amsthm}
\usepackage{amsmath}
\usepackage{amssymb}
\usepackage[unicode=true,pdfusetitle, bookmarks=true,bookmarksnumbered=false,bookmarksopen=false, breaklinks=false,pdfborder={0 0 0},backref=false,colorlinks=false] {hyperref}
\hypersetup{colorlinks,linkcolor=myurlcolor,citecolor=myurlcolor,urlcolor=myurlcolor}
\usepackage{graphics,epstopdf,graphicx}
\usepackage[up]{subfigure}
\usepackage{color}
\usepackage{times}
\usepackage{txfonts}
\usepackage{braket}
\usepackage{colortbl}
\definecolor{myurlcolor}{rgb}{0,0,0.7}
\def\dbar{{\mathchar'26\mkern-12mu d}}
\usepackage{cleveref}
\usepackage{verbatim}

\begin{document}

\title{Quantum R\'enyi Relative Entropies Affirm Universality of Thermodynamics}

\author{Avijit Misra}
\email{avijit@hri.res.in}
\affiliation{Harish-Chandra Research Institute, Allahabad, 211019, India}

\author{Uttam Singh}
\affiliation{Harish-Chandra Research Institute, Allahabad, 211019, India}

\author{Manabendra Nath Bera}
\affiliation{Harish-Chandra Research Institute, Allahabad, 211019, India}

\author{A. K. Rajagopal}
\email{attipat.rajagopal@gmail.com}
\affiliation{Inspire Institute Inc., Alexandria, Virginia, 22303, USA}
\affiliation{Harish-Chandra Research Institute, Allahabad, 211019, India}
\affiliation{ Institute of Mathematical Sciences, Chennai, 600113, India}

\begin{abstract}
We formulate a complete theory of quantum thermodynamics in the R\'enyi entropic formalism exploiting the R\'enyi relative entropies, starting from the maximum entropy principle. In establishing the first and second laws of quantum thermodynamics, we have \textit{correctly} identified accessible work and heat exchange both in equilibrium and non-equilibrium cases. The free energy (internal energy minus temperature times entropy) remains unaltered, when all the entities entering this relation are suitably defined. Exploiting R\'enyi relative entropies we have shown that this ``form invariance'' holds even beyond equilibrium and has profound operational significance in isothermal process. 
These results reduce to the Gibbs-von Neumann results when the R\'enyi entropic parameter  $\alpha$ approaches $1$. Moreover, it is shown that the \textit{universality} of the Carnot statement of the second law is the consequence of the form invariance of the free energy, which is in turn the consequence of maximum entropy principle. Further, the Clausius inequality, which is the precursor to the Carnot statement, is also shown to hold based on the data processing inequalities for the traditional and sandwiched R\'enyi relative entropies. Thus, we find that the thermodynamics of nonequilibrium state and its deviation from equilibrium together determine the thermodynamic laws. This is another important manifestation of the concepts of information theory in thermodynamics when they are extended to the quantum realm. Our work is a substantial step towards formulating a complete theory of quantum thermodynamics and corresponding resource theory.
\end{abstract}

\maketitle

\section{Introduction}
\label{intro}
The foundation of modern quantum thermodynamics research \cite{Gardas2015, Skrzypczyk2014, Vinjanampathya2014, Vinjanampathy2014, Binder2015} is based on von Neumann entropy. The maximum entropy principle \cite{JaynesA1957, JaynesB1957} with mean energy constraint gives rise to the Gibbs state which plays an important role in recent works on thermal operations and thermodynamical laws \cite{Brandao2015, Rudolph214, Rudolph114, Faist2015}. In this way the theory of quantum thermodynamics is developed based on von Neumann entropy. It is worth noting that the Gibbs states obtained consist of the exponential probability distributions. However, there are numerous physical systems which cannot be described by the Gibbsian exponential probability distribution and thereby inevitably needs the power law distributions \cite{Mandelbrot1982}. Precisely, the physics and thermodynamics of fractals and multifractals systems found in grain boundaries in metals, fluid dynamics, percolation, 
diffusion limited aggregation systems, DNA sequences are described suitably by using the R\'enyi entropy \cite{Mandelbrot1982, Mandelbrot1999, Jizba2004}.
Moreover, it is shown that R\'enyi entropy and its relative versions \cite{Tomamichel2015} are indispensable  in defining the second laws of quantum thermodynamics in microscopic regime \cite{Brandao2015, Rudolph214}.

It has been known for about half a century, R\'enyi entropy \cite{Renyi1961, Renyi1965} is endowed with all the necessary requirements for describing thermodynamics. 
Only recently \cite{Lenzi2000} the maximum entropy state with fixed energy was formulated and derived, giving rise to the R\'enyi thermal state indexed by the R\'enyi parameter, $\alpha$. When $\alpha\rightarrow1$, one obtains the von Neumann results. We construct here a complete theory of thermodynamics on par with the von Neumann theory. This generalization enlarges the scope of another facet of thermodynamics with several second laws \cite{Brandao2015} and also the Gibbs preserving maps giving rise to R\'enyi thermal state preserving maps \cite{Faist2015}. Another important outcome of this formalism is in establishing the \textit{universality} of second laws of thermodynamics stated as based on the Carnot statement. Furthermore, exploiting 
the data processing inequalities \cite{Tomamichel2015} obeyed by the two versions of R\'enyi relative entropies, the Clausius inequality is shown to hold. We thus find that the R\'enyi entropy and its relative versions  are the ingredients for establishing the second laws of thermodynamics.

The paper is organized as follows. In Sec. \ref{first-law}, we briefly introduce the thermodynamical quantities of interest in the R\'enyi formalism and the R\'enyi thermal states. Then we derive, following MaxEnt principle, the generalized first law of the R\'enyi thermodynamics. In Sec. \ref{second-law}, we develop and discuss the expression for free energy of non-equilibrium quantum states. Using the two versions of the R\'enyi relative entropy, we deduce the free energy for arbitrary quantum states and its  form invariant structure. In Sec. \ref{carnot}, we derive the efficiency of Carnot heat engine and thereby validate the universality of the second law of thermodynamics in the proposed form invariant formalism. Then the Clausius inequality, in the R\'enyi formalism, is considered in Sec. \ref{clausius}. Finally, in Sec. \ref{conclusion}, we outline a brief account of the results obtained and the possible implications thereof.  In the Appendices, details of finding Carnot efficiency and 
establishing the Clausius inequality in 
the R\'enyi formalism are given.

\smallskip{}

\section{Generalized first law of thermodynamics}
\label{first-law}

The Gibbs state, which is the equilibrium state, is obtained by maximizing the von Neumann entropy with a fixed internal energy. The maximum entropy (MaxEnt) principle \cite{JaynesA1957, JaynesB1957} is the underlying principle for such kind of equilibrium condition. It suggests that changing the definition of entropy functional, as well as the form of internal energy, gives rise to a new equilibrium state and hence a new theory of thermodynamics.

The R\'enyi entropy \cite{Renyi1965, Horodeckia1996}, which is a generalization of the von Neumann entropy, is given by $
 S_\alpha (\rho) = \frac{1}{1-\alpha} \ln (\mathrm{Tr}\rho^\alpha)$,
for a density matrix $\rho$ and $\alpha\in(0,1)\cup(1,\infty)$. The R\'enyi internal energy \cite{Lenzi2000} of $\rho$ is defined as $U_\alpha=\mathrm{Tr}[\rho^\alpha H]/\mathrm{Tr}\rho^\alpha$, where $H$ is the Hamiltonian of the system. Note that $S_\alpha (\rho)$ and $U_\alpha$ reduce to von Neumann entropy, $S(\rho)=-\mbox{Tr} \ (\rho \ln \rho)$ and internal or average energy $U=\mathrm{Tr}(\rho H)$ respectively, for $\alpha\rightarrow1$. The thermal equilibrium state for the R\'enyi entropy can be derived using MaxEnt principle, i.e., maximizing $S_\alpha(\rho)$ subject to a fixed internal energy $U_\alpha$, and is given by \cite{Lenzi2000}
\begin{align}
\label{equilibrium}
 \rho_{T\alpha} = \frac{1}{Z_\alpha} \left[1-(1-\alpha)\beta(H-U_{T\alpha})\right]^{1/(1-\alpha)}.
\end{align}
%
Here, $Z_\alpha = \mathrm{Tr}\left[\left\{1-(1-\alpha)\beta(H-U_{T\alpha})\right\}^{1/(1-\alpha)}\right]$ and $U_{T\alpha}=\mathrm{Tr}[\rho_{T\alpha}^\alpha H]/\mathrm{Tr}\rho_{T\alpha}^\alpha$.
The inverse temperature $\beta = 1/T$ (with the Boltzmann constant is set to unity and we follow this convention throughout the paper) is defined as
$\beta = \frac{\partial S_\alpha (\rho_{T\alpha})}{\partial U_{T\alpha} }$ which is a function of $\alpha$. Additionally the constraint $[1-(1-\alpha)\beta(H-U_{T\alpha}))]\geq 0$ is imposed to ensure the positive semi-definiteness of the thermal density matrix. Note that the R\'enyi thermal state reduces to the Gibbs state
when $\alpha\rightarrow 1$.  It should be noted that, similar to the Gibbs thermal state, one can also prepare R\'enyi thermal states, Eq. (\ref{equilibrium}), via environmental interaction and relaxation. A natural testbed for this would be multifractal systems among others \cite{Jizba2004}. Further, the equilibrium \textit{free energy} can be identified as $F_{T\alpha} = U_{T\alpha} -TS_\alpha(\rho_{T\alpha})$ \cite{Lenzi2000}. This general feature of the MaxEnt is independent of the choice of the form of the the density matrix \cite{Plastino1997}.

Now, we derive the generalized \textit{first law of thermodynamics} considering the change in equilibrium R\'enyi internal energy as
\begin{align}
\label{eq13}
dU_{T\alpha} = \frac{\mathrm{Tr}[d\rho_{T\alpha}^\alpha (H-U_{T\alpha})]}{\mathrm{Tr}\rho_{T\alpha}^\alpha} + \frac{\mathrm{Tr}(\rho_{T\alpha}^\alpha dH)}{\mathrm{Tr}\rho_{T\alpha}^\alpha}.
\end{align}
Under quasistatic isothermal process, the change in the entropy of the equilibrium state is $\beta \mathrm{Tr}[d\rho_{T\alpha}^\alpha (H-U_{T\alpha})]/\mathrm{Tr}\rho_{T\alpha}^\alpha$. Thus, the term $ \mathrm{Tr}[d\rho_{T\alpha}^\alpha (H-U_{T\alpha})]/\mathrm{Tr}\rho_{T\alpha}^\alpha$ can be identified as the heat exchanged. Moreover, $~\mathrm{Tr}(\rho_{T\alpha}^\alpha dH)/\mathrm{Tr}\rho_{T\alpha}^\alpha$ can be identified as the work done on the system, $~ \dbar W_{T\alpha}$, 
where it is considered to be the change in internal 
energy due to the change in an extensive parameter. Hence, the Eq. (\ref{eq13}) can be recast as
\begin{align}
\label{eq14}
 dU_{T\alpha}= \dbar Q_{T\alpha}  + \dbar W_{T\alpha}.
 \end{align}
This is the quantitative statement of the first law of thermodynamics following generalized theory of statistical mechanics based on the R\'enyi entropy. For a quasistatic isothermal process $~\dbar W_{T\alpha}=dF_{T\alpha}$, i.e., the infinitesimal change in the equilibrium free energy is the accessible work in the process. Therefore, for quasistatic isothermal processes we have
\begin{align}
 dU_{T\alpha}= dS_\alpha (\rho_{T\alpha})/\beta  + dF_{T\alpha}.
\end{align}
Note that the generalized first law of thermodynamics reduces to the well known first law of quantum thermodynamics (based on the von Neumann entropy) when $\alpha\rightarrow1.$

\smallskip{}

\section{Free energy for nonequilibrium states}
\label{second-law}
To this end we deal only with equilibrium thermodynamics. What if the system is away from equilibrium? In what follows, we study the thermodynamics of nonequilibrium states using the R\'enyi relative entropy with a motive to answer this question.

For a nonequilibrium quantum state $\rho_N$, which may be a solution of a dynamical master equation (such as in \cite{Rajagopal2002}), the R\'enyi entropy can be written as
  \begin{align}
S_\alpha (\rho_N)  = S_\alpha (\rho_{T\alpha}) - D_\alpha(\rho_N \parallel \rho_{T\alpha}) +\Delta_\alpha,
\label{RTET}
\end{align}
where $D_\alpha(\rho ~||~ \sigma)=\frac{1}{\alpha-1}\ln\mathrm{Tr}[\rho^\alpha\sigma^{1-\alpha}]$ is the ``traditional R\'enyi relative  entropy" between two quantum states $\rho$ and $\sigma$, and $\Delta_\alpha=\frac{1}{\alpha-1}\ln \left[ 1- \beta(1-\alpha)(U_{N\alpha}-U_{T\alpha}) \right]$ with $U_{N\alpha} = \mathrm{Tr}[\rho_N^\alpha H]/\mathrm{Tr}[\rho_N^\alpha]$ being the R\'enyi internal energy of $\rho_N$. Now we have 
\begin{align}
S_\alpha (\rho_N)= \beta\left[U_{N\alpha}- (F_{T\alpha} + \beta^{-1} D_\alpha(\rho_N \parallel \rho_{T\alpha}))  \right] + \Delta_\alpha',
\end{align}
where $\Delta_\alpha' = \left[\beta(U_{T\alpha}-U_{N\alpha})+\Delta_\alpha \right]$. One can easily check that $\Delta_\alpha' \rightarrow 0$ when $\alpha\rightarrow 1$ and the above equation reduces to the usual von Neumann case. 
Thus, for nonequilibrium states we have $ S_\alpha (\rho_N)= \beta(U_{N\alpha}- \tilde F_{N\alpha})$, where 
\begin{equation}
\tilde F_{N\alpha} = F_{T\alpha} + \beta^{-1} \left(D_\alpha(\rho_N \parallel \rho_{T\alpha})- \Delta_\alpha'\right) ,
\label{eq:modFE}
\end{equation}
is the modified free energy of the nonequilibrium state. Moreover, we show that $\tilde F_{N\alpha} \geqslant F_{T\alpha}$ (see \cref{Appx1}).

Considering again a quasistatic isothermal process, the change in entropy of the stationary nonequilibrium state is given by
\begin{align}
 dS_\alpha (\rho_N)&= \beta (dU_{N\alpha}- d\tilde F_{N\alpha})=\beta \left[~\dbar Q_{\alpha}-~\dbar Q_{hk\alpha}\right],
 \label{eq:dsn}
\end{align}
where $~\dbar Q_{\alpha}=  \mathrm{Tr}[d\rho_N^\alpha (H-U_{N\alpha})] / \mathrm{Tr}\rho_N^\alpha$, and $~\dbar Q_{hk\alpha}=d\tilde F_{N\alpha} -\dbar W_\alpha$, with ~$\dbar W_\alpha=\mathrm{Tr}(\rho_N^\alpha dH)/\mathrm{Tr}\rho_N^\alpha$. $~\dbar Q_{\alpha}$ is the total heat exchanged during the isothermal process and  $~\dbar Q_{hk\alpha}$ can be identified as the \textit{house-keeping} heat, in the same spirit as in stochastic thermodynamics \cite{Oono1998, Sasa2001, Seifert2012, Deffner2012, Gardas2015}, which is used to maintain the  nonequilibrium state away from thermal equilibrium. Now, for isothermal quasistatic processes in a generic quantum system, the Eq. (\ref{eq14}) can be recast as
\begin{align}
\label{eq:dun}
 d U_{N\alpha}& =~\dbar Q_{ex\alpha} +~ \dbar W_{ex\alpha} =\beta^{-1} d S_\alpha (\rho_N) + d\tilde F_{N\alpha}, 
\end{align}
where we denote excess heat as $~\dbar Q_{ex\alpha}=~\dbar Q_{\alpha}-~\dbar Q_{hk\alpha}$ and extractable work as $~~\dbar W_{ex\alpha}=\dbar Q_{hk\alpha}+\dbar W_\alpha=d\tilde F_{N\alpha}$. It is worth noticing that the notions of accessible work and the heat that results in entropy change, drastically differ from the equilibrium case (cf. Eq. (\ref{eq14})). Later, we will see that the notions of heat exchanged and accessible work in Eq. (\ref{eq14}) are not compatible in nonequilibrium scenario if the thermodynamical laws have to be valid in that case too.

The above analysis can also be carried by using the ``sandwiched  R\'enyi relative entropy" \cite{Tomamichel2013, Wilde2014}, which is defined as $
 \tilde D_\alpha(\rho \parallel \sigma) = \frac{1}{\alpha-1}\ln\mathrm{Tr}[\{\sigma^{\frac{1-\alpha}{2\alpha}} \rho \sigma^{\frac{1-\alpha}{2\alpha}}\}^{\alpha}]$,
between two quantum states $\rho$ and $\sigma$. As the sandwiched R\'enyi relative entropy incorporates the noncommutivity of the quantum states, unlike the traditional version, it is intuitively satisfactory to expect that the former can outperform the latter to unravel quantum features. Indeed, the profound advantages of the sandwiched relative entropy in studying classical capacity of a quantum channel~\cite{Wilde2014}, witnessing entanglement \cite{Rajagopal2014, Nayak2014}, quantum phase transitions \cite{Avijit2014}, nonasymptotic quantum information theory~\cite{Tomamichel2012,RennerPhD2005, Nilanjana2009}, etc,  have been noticed very recently.
Therefore, it is quite legitimate to extend the study of nonequilibrium quantum thermodynamics exploiting the sandwiched  R\'{e}nyi relative entropy.

The entropy of a nonequilibrium state $\rho_N$ can also be written
as 
\begin{align}
\label{eq20}
 S_\alpha (\rho_N) =  S_\alpha (\rho_{T\alpha})-\tilde D_\alpha(\rho_N \parallel \rho_{T\alpha})+\tilde\Delta_\alpha,
\end{align}
where $\tilde\Delta_\alpha=\frac{1}{\alpha-1}\ln \left[\mathrm{Tr} (A^{1/2\alpha}\rho A^{1/2\alpha})^\alpha/\mathrm{Tr}(\rho^\alpha) \right]$ and $A=\left[1-(1-\alpha)\beta(H-U_{T\alpha})\right]$. Thus, we have $ S_\alpha (\rho_N)= \beta(U_{N\alpha}- \mathcal{F}_{N\alpha})$, where
\begin{eqnarray}
\label{eq:smodfee}
\mathcal{F}_{N\alpha}=F_{T\alpha} + T \left(\tilde D_\alpha(\rho_N \parallel \rho_{T\alpha}) -\tilde\Delta_\alpha'\right),
\end{eqnarray}
is the modified free energy and $\tilde \Delta_\alpha' = \left[\beta(U_{T\alpha}-U_{N\alpha})+\tilde\Delta_\alpha \right]$. Again $\tilde \Delta_\alpha', \rightarrow 0$ when $\alpha\rightarrow 1$ and it recovers the von Neumann case. The nonequilibrium entropy change and the internal energy change for a quasistatic isothermal process can be derived following the same way as in the context of Eqs. (\ref{eq:dsn}) and (\ref{eq:dun}). Moreover, here also the change in the modified free energy $d\mathcal{F}_{N\alpha}$ can be distinguished as the accessible work in a quasistatic isothermal process and $\mathcal{F}_{N\alpha}\geq F_{T\alpha}$ (see \cref{Appx1}). 

\smallskip{}
\section{Second laws of thermodynamics based on Carnot statement}
\label{carnot}

\begin{figure}
 \centering
 \includegraphics[width=60 mm]{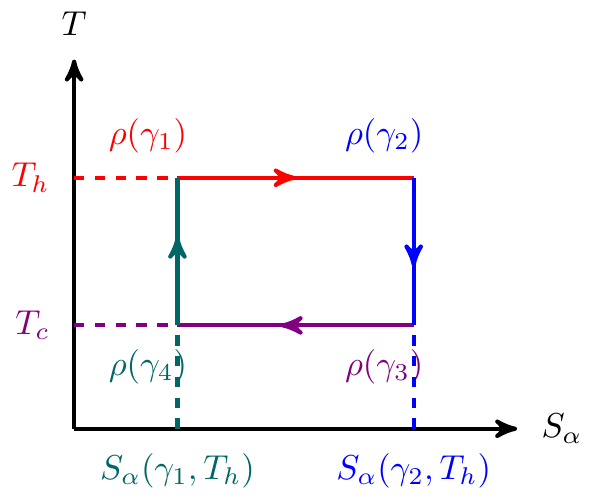}
 \caption{Schematic of Carnot cycle. $1\rightarrow2$ and $3\rightarrow4$ are the two isothermal steps at constant temperature $T_h$ and $T_c$ respectively. $2\rightarrow3$ and $4\rightarrow1$ are the adiabatic, isentropic steps. The different steps are performed by varying an external parameter $\gamma$.}\label{schematic}
\end{figure}

Now let us address the validity of second laws of thermodynamics based on the Carnot statement \cite{Callen1985, Gemmer2004} of the second law. 
%
Consider a four stroke Carnot engine operating between two reservoirs (heat baths), the hot and the cold with the temperatures $T_h$ and $T_c$ respectively (see Fig. \ref{schematic}). The baths consist of the R\'enyi thermal states with temperatures being $\alpha$ dependent. We find the efficiency  of this engine using the notions of accessible work and heat exchange which is responsible for entropy production developed earlier.
In the first step, the system absorbs $Q_{ex,1}=T_h[ S_\alpha(\gamma_2,T_h)- S_\alpha(\gamma_1,T_h)]$, amount of excess heat in a isothermal process at temperature $T_h$, from the the hot reservoir and the excess work $W_{ex,1}$, done by the system during the process is given by $\tilde F_{N\alpha}(\gamma_1,T_h)-\tilde F_{N\alpha}(\gamma_2,T_h)$, where $\gamma$ is the external parameter which is varied during the processes throughout the cycle. The system performs $W_{ex,2}$ work adiabatically in the second step and as it is an isentropic process the work done is at the cost of internal energy. As a consequence of performing work adiabatically, the temperature of the system falls down to $T_c$. Therefore in this step, $W_{ex,2}=\tilde F_{N\alpha}(\gamma_2,T_h)-\tilde F_{N\alpha}(\gamma_3,T_c)+(T_h-T_c) S_\alpha({\gamma_2},T_h)$. During the third step the work is actually done on the system in a isothermal process at temperature $T_c$ and the system releases some excess heat. The heat absorbed, $Q_{ex,3}$ and 
the work done, $W_{ex,3}$  by the system are given by $T_c[ S_\alpha(\gamma_4,T_c)- S_\alpha(\gamma_3,T_c)]$ and $\tilde F_{N\alpha}(\gamma_3,T_c)-\tilde F_{N\alpha}(\gamma_4,T_c)$, respectively. Note that we are always expressing the work done and heat absorbed by the system and we follow the same convention for all the four steps. In the fourth step, work is again performed on the system adiabatically.  As a result, the temperature of the system increases from $T_c$ to $T_h$ and the system is returned back to its initial state. The work done by the system is given by $W_{ex,4}=\tilde F_{N\alpha}(\gamma_4,T_c)-\tilde F_{N\alpha}(\gamma_1,T_h)-(T_h-T_c) S_\alpha(\gamma_1,T_h)$. Now, the Carnot efficiency which is the ratio of total work done by the system and the heat absorbed by the system in the first step,  is given by
\begin{align}
 \eta_C = \frac{W_{ex,1} + W_{ex,2} + W_{ex,3} + W_{ex,4}}{Q_{ex,1}} = 1-\frac{T_c}{T_h},
\end{align}
for arbitrary stationary quantum states.
%
Thus, the Carnot efficiency
matches with the one in classical thermodynamics. Importantly, the Carnot efficiency remains the same for both the traditional and sandwiched R\'enyi relative entropies. 
Since we
 consider arbitrary quantum states, it can be stated that the quantum correlation or coherence cannot be exploited to enhance the efficiency beyond the classical Carnot limit. 
 Therefore, if we account the accessible work and excess heat properly then ``{\it efficiency of any quantum engine undergoing a Carnot cycle is bounded above by $\eta_C$ }''. 
Thus, the Carnot statement of the second law of thermodynamics has been followed universally in the R\'enyi formulation, in parallel with the Gibbsian formulation of the same \cite{Gardas2015}. 

Note that the identification of heat exchange and accessible work in nonequilibrium scenario which results in accounting the accessible work by change in free energy (internal energy minus temperature times entropy) in isothermal processes, is consistent with the Carnot statement of the second law of thermodynamics. Thus, the form of free energy 
\begin{eqnarray}
\label{eq:genfee}
F_{\alpha}=U_{\alpha} - TS_{\alpha}
\end{eqnarray}
is valid (from operational viewpoint, like work extraction)  in nonequilibrium scenario too , where $T$ is the temperature of the corresponding heat bath (the relevant one, depending on the protocol). One may note that both the free energies in Eqs. (\ref{eq:modFE}) and (\ref{eq:smodfee}) are the same, i.e., $\mathcal{F}_{N\alpha}=\tilde F_{N\alpha}$, and given by the aforementioned form. It reflects that the form of the free energy of an nonequilibrium state is independent of the relative entropy ``distances", though the definition of free energy of the same is based on its ``distance" from the equilibrium one. A priori there is no reason why this form of free energy should be valid beyond equilibrium where the notion of temperature is not defined even. But this definition is consistent with second law of thermodynamics. Moreover, as free energy of the generalized thermal state is minimum among all the quantum states for all $\alpha$ and the change in free energy is the accessible work in an isothermal 
process, {\it it is not possible to extract work from a single heat bath}, which is another aspect of the second laws of thermodynamics. 
Thus, the apparent {\it universality} of second law of thermodynamics is a consequence of the {\it form invariance} of the free energy. Note this form of the free energy  
emerges naturally from the MaxEnt principle.

Interestingly, the apparently different forms of free energies discussed in \cite{Gardas2015} and \cite{ Skrzypczyk2014} are indeed equivalent and is a special case ($\alpha\rightarrow 1$) of the generalized free energy given in Eq. (\ref{eq:genfee}). In \cite{Skrzypczyk2014}, it is shown that if there exists any protocol by which one can extract more work than the free energy difference, then there would surely be a violation of the second law of thermodynamics. Similarly, in \cite{Gardas2015}, it is shown that the maximum extractable work, in any step in a Carnot cycle, has to be restricted by the free energy difference, if it has to be consistent with the second law 
of thermodynamics. Therefore, it is evident that the validity of the second laws of  thermodynamics is a consequence of the form invariance of the free energy when it is derived from the MaxEnt principle which sets the condition for equilibrium.

\smallskip{}

\section{Second laws based on Clausius inequality}
\label{clausius}
The second laws of the thermodynamics can further be substantiated in terms of  the Clausius inequality in the R\'enyi formalism. Consider nonequilibrium states which are close to the thermal equilibrium state, such that $U_{N\alpha}\approx U_{T\alpha}$, i.e., the difference of the  internal energies is small. Moreover, consider infinitesimal change of the nonequilibrium state $\rho_N$ by R\'enyi thermal state preserving map $\Upsilon$, i.e., $\rho_N \xrightarrow{\Upsilon} \rho_N + \delta \rho$,  where $\Upsilon$ is a completely positive trace preserving (CPTP) map that keeps the R\'enyi thermal state $\rho_{T\alpha}$ intact. As $\rho_N$ is close to $\rho_{T\alpha}$ and $\Upsilon$ introduces infinitesimal change in $\rho_N$, therefore, $\rho_N + \delta \rho$ is also close to $\rho_{T\alpha}$. Now, from Eq. (\ref{RTET}), the variation in the traditional R\'enyi relative entropy becomes
$ \delta D_\alpha(\rho_N \parallel \rho_{T\alpha})  = - \delta S_\alpha (\rho_N) +\delta \Delta_\alpha$,
where $ \delta D_\alpha(\rho_N \parallel \rho_{T\alpha})  =D_\alpha(\Upsilon[\rho_N] \parallel \Upsilon[\rho_{T\alpha}]) -  D_\alpha(\rho_N \parallel \rho_{T\alpha})$. Using the data processing inequality~\cite{Petz1986, Abe2003} for the traditional relative R\'enyi entropy which says that $\delta D_\alpha(\rho_N \parallel \rho_{T\alpha})\leq0$, for $\alpha \in [0,2]$,  we show that (see \cref{A1}),
\begin{align}
 \beta~ \delta Q_{total}\leq \delta S_\alpha(\rho_N).
 \label{cineq}
 \end{align}
This is nothing but the well known Clausius inequality. Also exploiting the data processing inequality for the sandwiched relative R\'enyi entropy for $\alpha \in [\frac{1}{2},\infty)$ \cite{Beigi2013, Frank2013, Tomamichel2013}, we show that the Clausius inequality holds for $\alpha\in[0,\infty)$, where $\rho_{T\alpha}, \ \rho_N$ and $\delta\rho$ are mutually commuting (see \cref{A1}).
Remarkably the Clausius inequality implies that the free energy is a monotone under R\'enyi thermal state preserving maps, when the Hamiltonian is kept fixed. This can also be seen from Eq. (\ref{eq:modFE}), when $U_{N\alpha}\approx U_{T\alpha}$. 

Now, we explore the Clausius inequality for noncommuting $\rho_{T\alpha},\rho_N$ and $\delta\rho$ in $\alpha > 2$. From Eq. (\ref{eq20}), we get
\begin{align}
\label{Eq:var-rel-ent}
 \delta \tilde D_\alpha(\rho_N \parallel \rho_{T\alpha}) = - \delta S_\alpha (\rho_N) + \beta \delta U_{N\alpha} +\delta \tilde\Delta'_\alpha,
\end{align}
where $\tilde\Delta'_\alpha=\frac{1}{\alpha-1}\ln \left[\mathrm{Tr} (A^{1/2\alpha}\rho A^{1/2\alpha})^\alpha/\mathrm{Tr}(\rho^\alpha) \right]- \beta(U_{N\alpha}-U_{T\alpha})$.
Using the data processing inequality for $\alpha \in [\frac{1}{2},\infty)$, we have
\begin{align}
\label{Eq:var-rel-ent-2}
 - \delta S_\alpha (\rho_N) + \beta \delta U_{N\alpha} +\delta \tilde\Delta'_\alpha \leq 0.
\end{align}
Therefore, if $\delta \tilde\Delta'_\alpha$ is either positive or vanishing to the first order then Clausius inequality holds for $\alpha \in [\frac{1}{2},\infty)$. For integer $\alpha$,  Eq. (\ref{Eq:var-rel-ent-2}) becomes an equality, to the first order and hence, if the Clausius inequality is satisfied then $\delta\tilde\Delta'_\alpha\geq0$ to the first order. We show that $\delta\tilde\Delta'_\alpha=0$ to the first order, with an explicit example (see \cref{ap:CIannex}). All these results indicate that for the Clausius inequality to hold in general, for any $\alpha$ and for any state, $ \delta\tilde\Delta'_\alpha$ has to be either positive or zero to the first order in the variation.
An analytical example considered in \cref{appendix:example} further supports the above observation.

\begin{figure}
\subfigure[~$\delta q$ = 0.001]{
\includegraphics[width=38 mm]{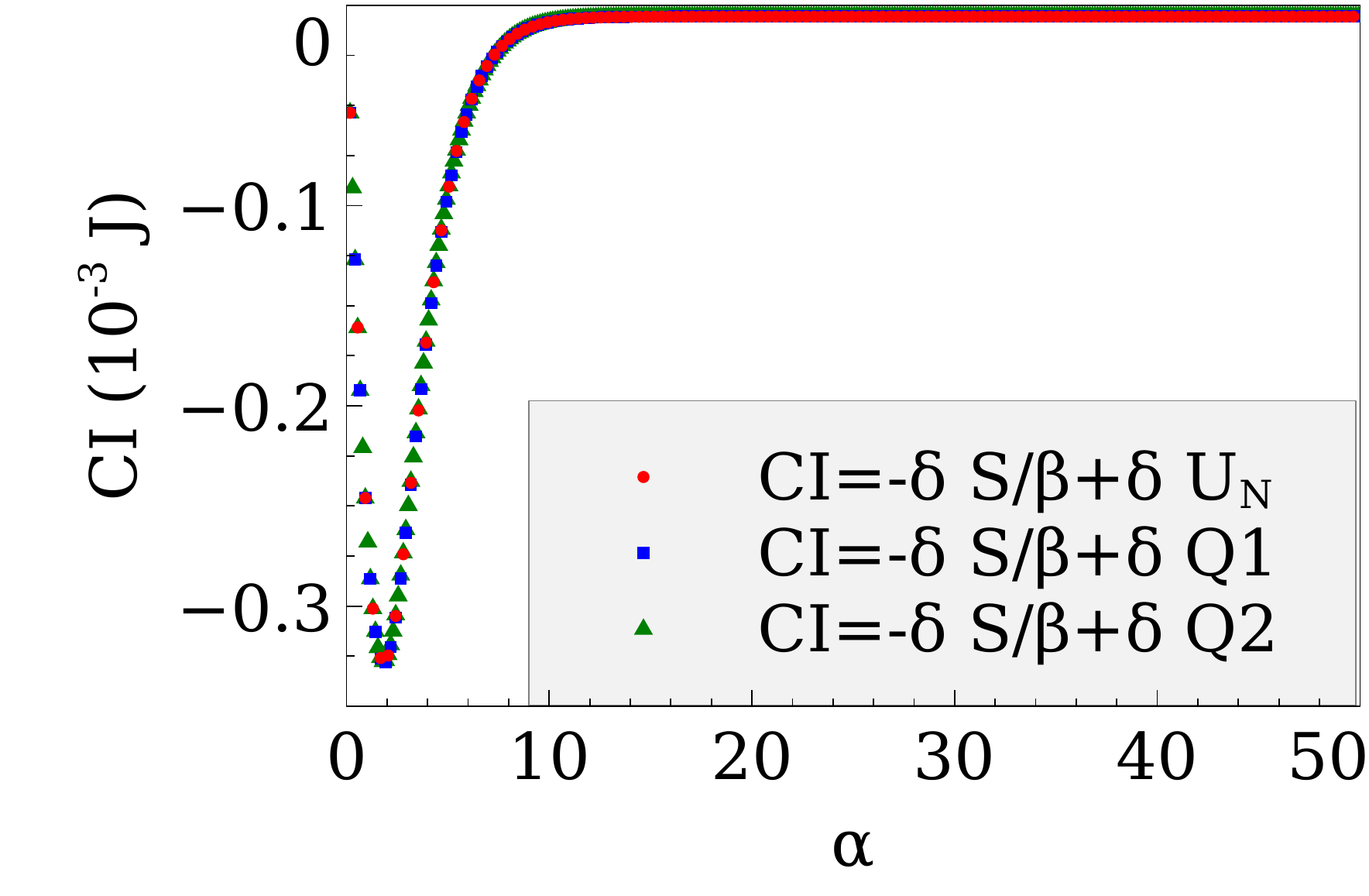}}
\subfigure[~$\delta q$ = 0.01]{
\includegraphics[width=38 mm]{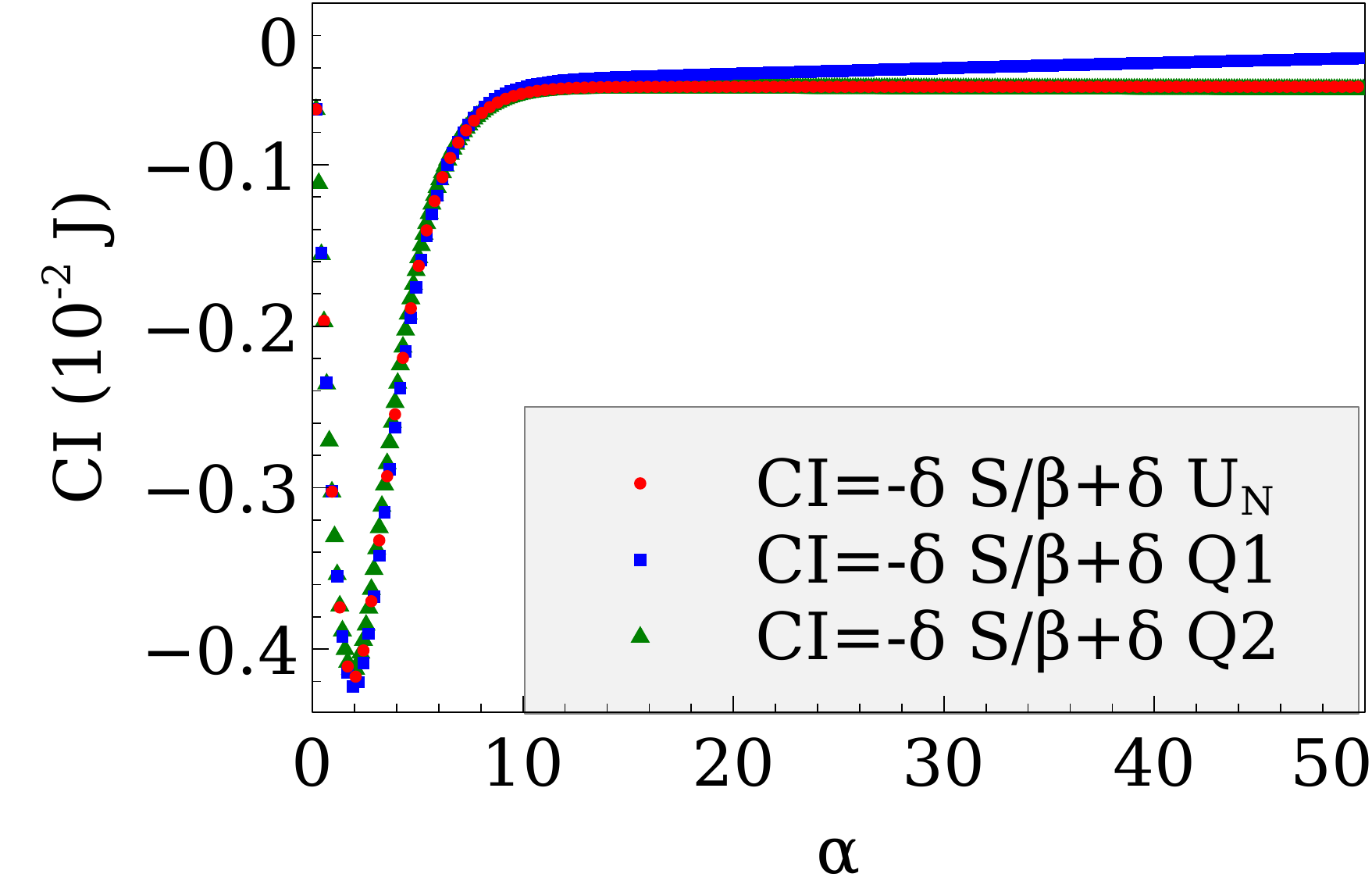}}
 \caption{(Color online) Clausius inequality for noncommuting quantum states for various values of $\alpha$: Along $y$-axis three different expressions for the Clausius inequality, which are the same to the first order, are plotted. The $x$ axis is dimensionless and $y$ axis has the dimension of energy (Joule).
Here, $ \delta Q_{1} = \mathrm{Tr}[d\rho_N^\alpha (H-U_{N\alpha})]/\mathrm{Tr}\rho_N^\alpha$ and $ \delta Q_{2} = \mathrm{Tr}[d\rho_N^\alpha (H-U_{T\alpha})]/\mathrm{Tr}\rho_N^\alpha$. We take $E_1=1$J, $p_0=0.7$ and $ \eta=  0.4 \ket{0}\bra{0} + 0.6 \ket{1}\bra{1} + 0.2 \ket{0}\bra{1} +0.2 \ket{1}\bra{0}$.
The different plots are for two different values of $\delta q$.}
 \label{fig1}
\end{figure}

To explore Clausius inequality numerically, we consider,
without loss of generality, the Hamiltonian of a qubit system to be $H=E_1\ket{1}\bra{1}$. The thermal state of the system is given by $ \rho_{T\alpha} = p_0 \ket{0}\bra{0} + (1-p_0)\ket{1}\bra{1}$ which fixes the inverse temperature as
 $\beta = (p_0^\alpha+p_1^\alpha)(p_0^{1-\alpha}- p_1^{1-\alpha})/E_1(1-\alpha)$. The nonequilibrium state is taken as $ \rho_N = (p_0 +\delta q)\ket{0}\bra{0} +((1-p_0)-\delta q)  \ket{1}\bra{1} +\delta q \ket{0}\bra{1}+\delta q \ket{1}\bra{0}$.
 The variation of $\rho_N$ is done by the R\'enyi thermal state preserving map $\Upsilon$ such that $\rho_N\xrightarrow{\Upsilon}\rho_N + \delta\rho =  (p_0+\delta q)\tau + (p_1-\delta q)\eta$, where $\eta$ is an arbitrary state and $\tau = (\rho_{T\alpha}-p_1\eta)/p_0$ \cite{Faist2015}. Fig. \ref{fig1}, indeed indicates that the Clausius inequality is respected, in general, for all ranges of $\alpha$.

Thus, we have the following results: the Clausius inequality holds
(i)  for $\alpha \in [0,2]$ which follows from the data processing inequality for the traditional R\'enyi case, and (ii) for $\alpha \in [\frac{1}{2},\infty)$ provided $ \delta\tilde\Delta'_\alpha$ is either positive or vanishing to the first order, which follows from the data processing inequality for the Sandwiched R\'enyi case . Moreover, the difference in the data processing inequalities \cite{Petz1986, Tomamichel2013, Beigi2013, Frank2013} for the two relative entropies is yet another manifestation of the different quantum features of the two, which is captured nicely in the context of Clausius inequality.

\vspace{0.05cm}

\smallskip{}
\section{Summary and concluding remarks}
\label{conclusion}
To formulate a complete theory of quantum thermodynamics based on the R\'enyi formalism, we  explicitly derive the generalized first law of thermodynamics, starting from the R\'enyi thermal state. The R\'enyi thermal state basically defines equilibrium temperature and hence the zeroth law of thermodynamics. The main thrust of this paper is to point out that the second laws of thermodynamics can be consistently derived based on the R\'enyi entropy and its relative versions. The overarching principle here is the well-known MaxEnt principle, given internal energy constraint.


Beyond equilibrium the notion of heat and work is not so well defined and whether the second law of thermodynamics is valid in this regime is an intriguing question.
We have successfully demonstrated that the proper identification of heat and work beyond equilibrium in fact plays the pivotal role to establish a valid second law which is respected universally. To correctly identify heat and work beyond equilibrium that is consistent with the second law of thermodynamics, we have exploited the equilibrium entropy associated with thermal density matrix and the relative entropy between an arbitrary density matrix and the thermal density matrix. Further, we have established that the \textit{universality} of the second law of thermodynamics based on the Carnot statement 
is the manifestation of this form invariance of free energy which holds even beyond equilibrium.

Using the data processing inequalities for traditional and sandwiched relative R\'enyi entropies we establish the validity of the Clausius inequality and hence strengthen the second laws of thermodynamics developed here.
 Our results, specifically, the monotonicity of free energy pave the way to an operational resource theory of quantum thermodynamics with the R\'enyi thermal state being the free state and the allowed operations being the R\'enyi thermal state preserving maps.  In this way, we have exhibited another important connection between the concepts of information theory and thermodynamics in modern quantum framework. 
 It may not be out of place here to point out the changes that the Renyi thermal state entails two known frameworks based on Gibbs thermal state: $(a)$ the Green function theory of many-particle systems \cite{Rajagopal1998, LenziB2000, Martin1959} and hence also $(b)$ the Kubo-Martin-Schwinger (KMS) condition \cite{Martin1959}.
 \smallskip{}

\noindent{\emph{Acknowledgments}}---The authors thank A. R. Usha Devi and M. M. Wilde for useful remarks on the manuscript. A.M., U.S. and M.N.B. acknowledge the research fellowship of Department of Atomic Energy, Government of India.

\appendix
\setcounter{equation}{0}

\section{Free energy is minimum for thermal states}
\label{Appx1}
The free energy of an arbitrary quantum state is larger than that of a thermal equilibrium state, i.e., $ \tilde F_{N\alpha} > F_{T\alpha}$, for any $\alpha$. This follows from Eq.
\begin{eqnarray}
\mathcal{F}_{N\alpha}=F_{T\alpha} + T \left(\tilde D_\alpha(\rho_N \parallel \rho_T) -\tilde\Delta_\alpha'\right),
\end{eqnarray}
$D_\alpha(\rho_N \parallel \rho_T)\geqslant 0$ and $\Delta_\alpha' \leqslant 0$. The first inequality, $D_\alpha(\rho_N \parallel \rho_T)\geqslant 0$, is due to the positivity of the R\'enyi relative entropy. The negativity of the latter quantity, can be shown by demanding the condition $\beta(\alpha-1)(U_{N\alpha}-U_{T\alpha})>-1$, which is the cut off condition of consistent probabilistic interpretation of R\'enyi thermal state, and the inequality $\ln(1+x)\leq x$ for all $x>-1$. Thus, confirming the known result that MaxEnt state implies minimum free energy.

\section{Proof of the Clausius inequality in \texorpdfstring{$\alpha \in [0,2] $ }{} using traditional R\'enyi relative entropy}
\label{A1}
The R\'enyi relative entropy between a nonequilibrium state $\rho_N$ and the thermal equilibrium state $\rho_T$, is given by
\begin{align}
 D_\alpha(\rho_N \parallel \rho_T)  = S_\alpha (\rho_T) - S_\alpha (\rho_N) +\Delta_\alpha,
\end{align}
where $\Delta_\alpha=\frac{1}{\alpha-1}\ln \left[ 1- \beta(1-\alpha)(U_{N\alpha}-U_{T\alpha}) \right]$. Consider an infinitesimal change in the density matrix of the system, $\rho_N \rightarrow \Upsilon[\rho_N] = \rho_N + \delta \rho$ via a CPTP map that keeps the R\'enyi thermal state intact. We dub such maps as R\'enyi thermal state preserving maps. The change in the traditional R\'enyi relative entropy under such maps, is given by
\begin{align}
\label{trad-delta}
 \delta D_\alpha(\rho_N \parallel \rho_T)  &= D_\alpha(\Upsilon[\rho_N] \parallel \rho_T) - D_\alpha(\rho_N \parallel \rho_T) \nonumber\\
 &=  - \delta S_\alpha (\rho_N) +  \delta\Delta_\alpha,
\end{align}
where
\begin{align}
\label{delta-heat}
 \delta \Delta_\alpha &=\frac{- \beta(1-\alpha)}{(\alpha-1)\left[ 1- \beta(1-\alpha)(U_{N\alpha}-U_{T\alpha}) \right]}   \delta (U_{N\alpha} - U_{T\alpha}) \nonumber\\
 &=\frac{- \beta(1-\alpha)}{(\alpha-1)\left[ 1- \beta(1-\alpha)(U_{N\alpha}-U_{T\alpha}) \right]}   \delta U_{N\alpha}\nonumber\\
 &= \frac{ \beta}{\left[ 1- \beta(1-\alpha)(U_{N\alpha}-U_{T\alpha}) \right]} \left(\frac{\mathrm{Tr}[ \delta\rho_N^\alpha (H - U_{N\alpha})]}{\mathrm{Tr}(\rho_N^\alpha)} \right)\nonumber\\
&=\beta~ \delta Q_{total}(1+\beta(1-\alpha)(U_{N\alpha}-U_{T\alpha})).\\
&\approx \beta~ \delta Q_{total},
\end{align}
 We have used the fact that $\delta Q_{total}=\delta U_N$,  as Hamiltonian $H$ remains unchanged and $(U_{N\alpha}-U_{T\alpha})$ is very small.
Thus, we have
\begin{align}
\label{clausiusappen}
  \delta D_\alpha(\rho_N \parallel \rho_T)  = - \delta S_\alpha (\rho_N) + \beta~ \delta Q_{total}.
\end{align}
Now, using the data processing inequality for traditional R\'enyi relative entropy \cite{Petz1986, Abe2003}, $D_\alpha(\Upsilon[\rho_N] \parallel \Upsilon[\rho_T]) \leq D_\alpha(\rho_N \parallel \rho_T)$, for $\alpha \in [0,2]$, we have
\begin{align}
 - \delta S_\alpha (\rho_N) + \beta~ \delta Q_{total} \leq 0.
\end{align}
Therefore, we have $\delta S_\alpha (\rho_N)  \geq  \beta~ \delta Q_{total}$ for $\alpha \in [0,2]$, which is a statement of the second law of thermodynamics in terms of Clausius inequality.
The Clausius inequality for transformations under unital maps near the equilibrium was shown in \cite{Abe2003} for $\alpha \in (0,2]$, by a different approach. However, the Clausius inequality derived above applies to the R\'enyi thermal state preserving operations.
Clausius inequality has also been shown in \cite{Vinjanampathya2014, Vinjanampathy2014, Binder2015} for von Neumann case. 

Notice that if $\rho_T, \ \rho_N$ and $\delta\rho$ are mutually commuting then the traditional relative entropies can be replaced by the sandwiched ones in Eq. (\ref{clausiusappen}) and following the data processing inequality for sandwiched relative entropies \cite{Tomamichel2013, Wilde2014} which holds for $\alpha \in [\frac{1}{2},\infty)$, the Clausius inequality can be established for $\alpha \in [0,\infty)$ for commuting case.

\section{Clausius inequality for noncommuting states \label{ap:CIannex}}

 If  $\rho_N +\delta \rho$, $\rho_N$ and $\rho_T$ are mutually commuting, then both $\tilde D_\alpha(\rho_N+\delta\rho||\rho_T)$ and $\tilde D_\alpha(\rho_N||\rho_T)$ vanish independently to the first order for our case of interest, i.e., for close by $\rho_N$ and $\rho_T$ and for $\alpha\in[\frac{1}{2},\infty)$, we have
\begin{align}
\label{RSETM}
 \delta \tilde D_\alpha(\rho_N \parallel \rho_T)  =  -\delta S_{\alpha} +\beta ~\delta U_{N\alpha} + \delta\tilde\Delta'_{\alpha}\approx0,
\end{align}
where $\delta\tilde\Delta'_{\alpha}$ is the variation of $\tilde\Delta'_\alpha=\frac{1}{\alpha-1}\ln \left[\mathrm{Tr} (A^{1/2\alpha}\rho_N A^{1/2\alpha})^\alpha/\mathrm{Tr}(\rho_N^\alpha) \right]-\beta(U_{N\alpha}-U_{ T\alpha})$. In this case it can be shown that $\delta\tilde\Delta'_{\alpha} \approx 0$. From Eq. (\ref{RSETM}), to the first order,  we have
\begin{align}
 -\delta S_{\alpha} +\beta ~\delta U_{N\alpha} \approx 0.
\end{align}
Similarly, even if the states $\rho_N +\delta \rho$, $\rho_N$ and $\rho_T$ are not mutually commuting, it can easily be shown that for the integer $\alpha$, the Sandwiched and traditional R\'enyi relative entropies vanish independently to first order. In this case, if the Clausius inequality is to be satisfied, it amounts to requiring that $\delta\tilde\Delta'_{\alpha} \geq 0$ to first order in variation. Further, we consider an example of noncommuting states to supplement our observations.

\section{Clausius inequality for an analytical example}
\label{appendix:example}

Without loss of generality, the Hamiltonian of the system can be considered as $H=E_1\ket{1}\bra{1}$ \cite{Fernandoa2013}. Let the thermal state be given by
\begin{align}
\label{thermal-state}
 \rho_T =
 \begin{pmatrix}
  p_0 & 0\\
  0 & p_1
 \end{pmatrix},
 \end{align}
 with $p_0+p_1=1$. This fixes the inverse temperature as
 \begin{align}
 \label{beta}
  \beta = \frac{p_0^\alpha+p_1^\alpha}{E_1(1-\alpha)}(p_0^{1-\alpha}- p_1^{1-\alpha}).
 \end{align}
 Consider a nonequilibrium state, which is close to the thermal state, as
 \begin{align}
 \label{athermal-state}
 \rho_N = (p_0 +2\delta q)\ket{e_0}\bra{e_0} + (p_1-2\delta q)\ket{e_1}\bra{e_1},
 \end{align}
 where $ \ket{e_0} = \frac{1}{\sqrt{1+2\delta q}}( 1+\delta q ,  \delta q  )^T$ and $ \ket{e_1} = \frac{1}{\sqrt{1+2\delta q}}( -\delta q,  1+\delta q)^T$.
 Also, consider a small variation of $\rho_N$, given by
 \begin{align}
  \label{athermal-state-plus-variation}
 \rho_N + \delta\rho =  (p_0 +\delta q)\ket{e_0}\bra{e_0} + (p_1-\delta q)\ket{e_1}\bra{e_1},
 \end{align}
 where  $ \delta\rho =  -\delta q\ket{e_0}\bra{e_0} +\delta q\ket{e_1}\bra{e_1} $.  Note that $\rho_T$ does not commute with $\rho_N$ and $\rho_N+\delta \rho$. For this example, we have shown that the Clausius inequality is satisfied as an equality to the first order in $\delta q$ and furthermore, it is shown that $\delta\tilde\Delta'_\alpha$ is indeed zero.

As a consequence of the variation $\rho_N \rightarrow \rho_N + \delta \rho$, the change in the entropy of the system is given by
 \begin{align}
 \label{entropy-change}
 - \delta S_\alpha &= -[S_\alpha(\rho_N+\delta\rho) - S_\alpha(\rho_N)]\nonumber\\
 &=-\frac{1}{1-\alpha}[\ln\{\mathrm{Tr}[(\rho_N+\delta\rho)^\alpha]\} - \ln\{\mathrm{Tr}[\rho_N^\alpha]\} ] \nonumber\\
 &=-\frac{1}{1-\alpha}[\ln\{(p_0 +\delta q)^\alpha + (p_1 - \delta q)^\alpha\}\nonumber\\
 & ~~~~~~~~~~~~~~~~~~~~- \ln\{(p_0 + 2\delta q)^\alpha + (p_1 - 2\delta q)^\alpha\} ] \nonumber\\
 &=\frac{\alpha~\delta q}{1-\alpha}\left[\frac{p_0^{\alpha-1} - p_1^{\alpha-1}}{p_0^\alpha + p_1^\alpha}\right].
 \end{align}
 Since the nonequilibrium state is close to the thermal state, the change in the internal energy is same as change in heat to the first order in variation, i.e. $\delta Q_{\alpha} \approx  U_{N\alpha}(\rho_N+\delta\rho)-U_{N\alpha}(\rho_N)$, therefore, we have
 \begin{align}
 \label{heat-change}
\delta Q_{\alpha}& \approx \frac{ \mathrm{Tr}[(\rho_N +\delta\rho)^\alpha H] }{\mathrm{Tr}[(\rho_N +\delta\rho)^\alpha]} - \frac{ \mathrm{Tr}[\rho_N^\alpha H] }{\mathrm{Tr}[\rho_N^\alpha]}\nonumber\\
 &=E_1\frac{ (p_0+\delta q)^\alpha |\bra{e_0}1\rangle|^2 + (p_1-\delta q)^\alpha |\bra{e_1}1\rangle|^2}{(p_0+\delta q)^\alpha+   (p_1-\delta q)^\alpha}\nonumber\\ 
 &~~-E_1\frac{ (p_0+2\delta q)^\alpha |\bra{e_0}1\rangle|^2 + (p_1-2\delta q)^\alpha |\bra{e_1}1\rangle|^2}{(p_0+2\delta q)^\alpha+   (p_1-2\delta q)^\alpha}\nonumber\\
 &= \alpha~ \delta q~ E_1\frac{ p_1^\alpha (p_0^{\alpha-1 }- p_1^{\alpha-1 }) +p_1^{\alpha-1 } (p_0^\alpha + p_1^\alpha) }{(p_0^\alpha + p_1^\alpha)^2 } \nonumber\\
 &= \alpha~ \delta q~ E_1\frac{ (p_0p_1)^{\alpha-1 }}{(p_0^\alpha + p_1^\alpha)^2 }.
 \end{align}
 Now, using Eq. (\ref{beta}), we have
 \begin{align}
\beta ~\delta Q_{\alpha}& =  \frac{\alpha~ \delta q}{1-\alpha}(p_0^{1-\alpha}- p_1^{1-\alpha})\frac{ (p_0p_1)^{\alpha-1 }}{(p_0^\alpha + p_1^\alpha) } \nonumber\\
&=-\frac{\alpha~\delta q}{1-\alpha}\left[\frac{p_0^{\alpha-1} - p_1^{\alpha-1}}{p_0^\alpha + p_1^\alpha}\right].
 \end{align}
Combining Eq. (\ref{entropy-change}) and Eq. (\ref{heat-change}), we get $- \delta S_\alpha + \beta~ \delta Q_{\alpha} =0$. Therefore, the Clausius inequality is satisfied for all values of $\alpha$, for the qubit case considered above. Moreover, for this case, the variation in the sandwiched R\'enyi relative entropy can be shown to be zero to first order. 
The change in sandwiched R\'enyi relative entropy is given by 
 \begin{align}
 (1-\alpha)~\delta\tilde D_\alpha(\rho_N ||\rho_T) &=   \ln \mathrm{Tr}\left[(A+B)^\alpha\right]-  \ln \mathrm{Tr}\left[A^\alpha\right],
\end{align}
 where
\begin{align}
&A= \rho_T^{\frac{1-\alpha}{2\alpha}} (\rho_N+\delta\rho)~ \rho_T^{\frac{1-\alpha}{2\alpha}} = \sum_{i,j} a_{ij} \ket{i}\bra{j};\\
&B= \rho_T^{\frac{1-\alpha}{2\alpha}}(-\delta\rho)~ \rho_T^{\frac{1-\alpha}{2\alpha}} =\delta q\sum_{i,j} b_{ij} \ket{i}\bra{j},
\end{align}
with
$
a_{ij} = (p_i p_j)^{\frac{1-\alpha}{2\alpha}}\sum_k q_k \langle i \ket{e_k}\bra{e_k}j \rangle$,
$q_0=p_0+\delta q$ and $q_1=p_1-\delta q$ and $
b_{ij} = (p_i p_j)^{\frac{1-\alpha}{2\alpha}}\bra{i}(|e_0\rangle\langle e_0| - |e_1\rangle\langle e_1|) \ket{j}$.
To the first order, we have $a_{00} = p_0^{\frac{1-\alpha}{\alpha}} (p_0+\delta q)$, $ a_{11} = p_1^{\frac{1-\alpha}{\alpha}} (p_1-\delta q)$, $ a_{01} = (p_0p_1)^{\frac{1-\alpha}{2\alpha}} \delta q(p_0 -p_1+2\delta q) = a_{10}$, $ b_{00} = p_0^{\frac{1-\alpha}{\alpha}}$,  $ b_{11} = -p_1^{\frac{1-\alpha}{\alpha}}$, $b_{01}=0=b_{10}$. The eigenvalues of $A$ are given by
\begin{align}
\label{muss}
\mu(s) &= \frac{a_{00} + a_{11} + s \sqrt{(a_{00}-a_{11})^2 + 4|a_{01}|^2}}{2}\nonumber\\
&\approx \frac{a_{00} + a_{11} + s ~|a_{00}-a_{11}|}{2},
\end{align}
where $s=\{1,-1\}$. Similarly, the eigenvalues of $(A+B)$ are given by
\begin{align}
\nu(s) = \frac{c_{00} + c_{11} + s \sqrt{(c_{00}-c_{11})^2 + 4|c_{01}|^2}}{2};~~~~\{s=1,-1\},
\end{align}
where $c_{ij} = a_{ij} + \delta q~ b_{ij}$. To the first order, we have
\begin{align}
&\sqrt{ (c_{00} - c_{11})^2 + 4|c_{01}|^2}\nonumber\\
&\approx |a_{00}-a_{11}| + \delta q \frac{ (a_{00} - a_{11})(b_{00} - b_{11})}{|a_{00}-a_{11}|}.
\end{align}
Therefore, $ \nu(s) =\mu(s)+ \delta q~ m(s)$,
where $\mu(s)$ is given by Eq. (\ref{muss}) and
\begin{align}
 &m(s) = \frac{b_{00} +b_{11}+s  \frac{ (a_{00} - a_{11})(b_{00} - b_{11})}{|a_{00}-a_{11}|}}{2}.
\end{align}
For $a_{00} \geq a_{11}$, we have $\mu(1) = a_{00}$, $\mu(-1) = a_{11}$,  $m(1) = b_{00}$,  $m(-1) = b_{11}$. For $a_{00} \leq a_{11}$, we have $\mu(1) = a_{11}$, $\mu(-1) = a_{00}$,  $m(1) = b_{11}$,  $m(-1) = b_{00}$.
Now,
\begin{align}
 &\ln\mathrm{Tr}[(A+B)^\alpha] = \ln\sum_{s=\{-1,1\}} \mu(s)^\alpha + \alpha\delta q_0~ \mu(s)^{\alpha-1} m(s)\nonumber\\
 &= \ln[ \mu(1)^\alpha + \mu(-1)^\alpha]
 + \alpha~\delta q \frac{\mu(1)^{\alpha-1} m(1) + \mu(-1)^{\alpha-1} m(-1)}{[ \mu(1)^\alpha + \mu(-1)^\alpha]}.\nonumber
 \end{align}
 Also, we have $\ln\mathrm{Tr}[A^\alpha] = \ln[ \mu(1)^\alpha + \mu(-1)^\alpha]$. Therefore,
 \begin{align}
  (1-\alpha) \delta\tilde D_\alpha(\rho_N ||\rho_T)  &= \alpha~\delta q \left(\frac{\mu(1)^{\alpha-1} m(1) + \mu(-1)^{\alpha-1} m(-1)}{\mu(1)^\alpha + \mu(-1)^\alpha}\right)\nonumber\\
  &\approx \alpha~\delta q_0~ \left(\frac{a_{00}^{\alpha-1}b_{00} + a_{11}^{\alpha-1} b_{11}}{ a_{00}^\alpha +a_{11}^\alpha}\right)\nonumber\\
   &\approx \alpha~\delta q_0~ \left(\frac{p_0^{\frac{\alpha-1}{\alpha}} p_0^{\frac{1-\alpha}{\alpha}} - p_1^{\frac{\alpha-1}{\alpha}} p_1^{\frac{1-\alpha}{\alpha}} }{ p_0+p_1}\right)\nonumber\\ 
   &\approx 0,
 \end{align}
 where in the second line we have used values of $\mu(s)$ and $m(s)$ for the case $a_{00}\geq a_{11}$. It can be seen easily that the second line is the same for the case $a_{00}\geq a_{11}$, too.
Therefore, $\delta\tilde D_\alpha(\rho_N ||\rho_T) \approx 0$. Hence, for this case, we have $\delta\tilde\Delta'_\alpha$ vanishing to first order, thereby supporting our observation. 
 
\bibliographystyle{apsrev4-1}
\bibliography{renyi-lit}

\begin{thebibliography}{43}%
\makeatletter
\providecommand \@ifxundefined [1]{%
 \@ifx{#1\undefined}
}%
\providecommand \@ifnum [1]{%
 \ifnum #1\expandafter \@firstoftwo
 \else \expandafter \@secondoftwo
 \fi
}%
\providecommand \@ifx [1]{%
 \ifx #1\expandafter \@firstoftwo
 \else \expandafter \@secondoftwo
 \fi
}%
\providecommand \natexlab [1]{#1}%
\providecommand \enquote  [1]{``#1''}%
\providecommand \bibnamefont  [1]{#1}%
\providecommand \bibfnamefont [1]{#1}%
\providecommand \citenamefont [1]{#1}%
\providecommand \href@noop [0]{\@secondoftwo}%
\providecommand \href [0]{\begingroup \@sanitize@url \@href}%
\providecommand \@href[1]{\@@startlink{#1}\@@href}%
\providecommand \@@href[1]{\endgroup#1\@@endlink}%
\providecommand \@sanitize@url [0]{\catcode `\\12\catcode `\$12\catcode
  `\&12\catcode `\#12\catcode `\^12\catcode `\_12\catcode `\%12\relax}%
\providecommand \@@startlink[1]{}%
\providecommand \@@endlink[0]{}%
\providecommand \url  [0]{\begingroup\@sanitize@url \@url }%
\providecommand \@url [1]{\endgroup\@href {#1}{\urlprefix }}%
\providecommand \urlprefix  [0]{URL }%
\providecommand \Eprint [0]{\href }%
\providecommand \doibase [0]{http://dx.doi.org/}%
\providecommand \selectlanguage [0]{\@gobble}%
\providecommand \bibinfo  [0]{\@secondoftwo}%
\providecommand \bibfield  [0]{\@secondoftwo}%
\providecommand \translation [1]{[#1]}%
\providecommand \BibitemOpen [0]{}%
\providecommand \bibitemStop [0]{}%
\providecommand \bibitemNoStop [0]{.\EOS\space}%
\providecommand \EOS [0]{\spacefactor3000\relax}%
\providecommand \BibitemShut  [1]{\csname bibitem#1\endcsname}%
\let\auto@bib@innerbib\@empty
\bibitem [{\citenamefont {{Gardas}}\ and\ \citenamefont
  {{Deffner}}()}]{Gardas2015}%
  \BibitemOpen
  \bibfield  {author} {\bibinfo {author} {\bibfnamefont {B.}~\bibnamefont
  {{Gardas}}}\ and\ \bibinfo {author} {\bibfnamefont {S.}~\bibnamefont
  {{Deffner}}},\ }\href@noop {} {}\Eprint {http://arxiv.org/abs/1503.03455}
  {arXiv:1503.03455} \BibitemShut {NoStop}%
\bibitem [{\citenamefont {Skrzypczyk}\ \emph {et~al.}(2014)\citenamefont
  {Skrzypczyk}, \citenamefont {Short},\ and\ \citenamefont
  {Popescu}}]{Skrzypczyk2014}%
  \BibitemOpen
  \bibfield  {author} {\bibinfo {author} {\bibfnamefont {P.}~\bibnamefont
  {Skrzypczyk}}, \bibinfo {author} {\bibfnamefont {A.~J.}\ \bibnamefont
  {Short}}, \ and\ \bibinfo {author} {\bibfnamefont {S.}~\bibnamefont
  {Popescu}},\ }\href {http://dx.doi.org/10.1038/ncomms5185} {\bibfield
  {journal} {\bibinfo  {journal} {Nature Communications}\ }\textbf {\bibinfo
  {volume} {5}} (\bibinfo {year} {2014})}\BibitemShut {NoStop}%
\bibitem [{\citenamefont {Vinjanampathy}\ and\ \citenamefont
  {Modi}()}]{Vinjanampathya2014}%
  \BibitemOpen
  \bibfield  {author} {\bibinfo {author} {\bibfnamefont {S.}~\bibnamefont
  {Vinjanampathy}}\ and\ \bibinfo {author} {\bibfnamefont {K.}~\bibnamefont
  {Modi}},\ }\href@noop {} {}\Eprint {http://arxiv.org/abs/1411.7755}
  {arXiv:1411.7755} \BibitemShut {NoStop}%
\bibitem [{\citenamefont {{Vinjanampathy}}\ and\ \citenamefont
  {{Modi}}()}]{Vinjanampathy2014}%
  \BibitemOpen
  \bibfield  {author} {\bibinfo {author} {\bibfnamefont {S.}~\bibnamefont
  {{Vinjanampathy}}}\ and\ \bibinfo {author} {\bibfnamefont {K.}~\bibnamefont
  {{Modi}}},\ }\href@noop {} {}\Eprint {http://arxiv.org/abs/1405.6140}
  {arXiv:1405.6140} \BibitemShut {NoStop}%
\bibitem [{\citenamefont {Binder}\ \emph {et~al.}(2015)\citenamefont {Binder},
  \citenamefont {Vinjanampathy}, \citenamefont {Modi},\ and\ \citenamefont
  {Goold}}]{Binder2015}%
  \BibitemOpen
  \bibfield  {author} {\bibinfo {author} {\bibfnamefont {F.}~\bibnamefont
  {Binder}}, \bibinfo {author} {\bibfnamefont {S.}~\bibnamefont
  {Vinjanampathy}}, \bibinfo {author} {\bibfnamefont {K.}~\bibnamefont {Modi}},
  \ and\ \bibinfo {author} {\bibfnamefont {J.}~\bibnamefont {Goold}},\ }\href
  {\doibase 10.1103/PhysRevE.91.032119} {\bibfield  {journal} {\bibinfo
  {journal} {Phys. Rev. E}\ }\textbf {\bibinfo {volume} {91}},\ \bibinfo
  {pages} {032119} (\bibinfo {year} {2015})}\BibitemShut {NoStop}%
\bibitem [{\citenamefont {Jaynes}(1957{\natexlab{a}})}]{JaynesA1957}%
  \BibitemOpen
  \bibfield  {author} {\bibinfo {author} {\bibfnamefont {E.~T.}\ \bibnamefont
  {Jaynes}},\ }\href {\doibase 10.1103/PhysRev.106.620} {\bibfield  {journal}
  {\bibinfo  {journal} {Phys. Rev.}\ }\textbf {\bibinfo {volume} {106}},\
  \bibinfo {pages} {620} (\bibinfo {year} {1957}{\natexlab{a}})}\BibitemShut
  {NoStop}%
\bibitem [{\citenamefont {Jaynes}(1957{\natexlab{b}})}]{JaynesB1957}%
  \BibitemOpen
  \bibfield  {author} {\bibinfo {author} {\bibfnamefont {E.~T.}\ \bibnamefont
  {Jaynes}},\ }\href {\doibase 10.1103/PhysRev.108.171} {\bibfield  {journal}
  {\bibinfo  {journal} {Phys. Rev.}\ }\textbf {\bibinfo {volume} {108}},\
  \bibinfo {pages} {171} (\bibinfo {year} {1957}{\natexlab{b}})}\BibitemShut
  {NoStop}%
\bibitem [{\citenamefont {Brand\~ao}\ \emph {et~al.}(2015)\citenamefont
  {Brand\~ao}, \citenamefont {Horodecki}, \citenamefont {Ng}, \citenamefont
  {Oppenheim},\ and\ \citenamefont {Wehner}}]{Brandao2015}%
  \BibitemOpen
  \bibfield  {author} {\bibinfo {author} {\bibfnamefont {F.}~\bibnamefont
  {Brand\~ao}}, \bibinfo {author} {\bibfnamefont {M.}~\bibnamefont
  {Horodecki}}, \bibinfo {author} {\bibfnamefont {N.}~\bibnamefont {Ng}},
  \bibinfo {author} {\bibfnamefont {J.}~\bibnamefont {Oppenheim}}, \ and\
  \bibinfo {author} {\bibfnamefont {S.}~\bibnamefont {Wehner}},\ }\href
  {\doibase 10.1073/pnas.1411728112} {\bibfield  {journal} {\bibinfo  {journal}
  {Proceedings of the National Academy of Sciences}\ }\textbf {\bibinfo
  {volume} {112}},\ \bibinfo {pages} {3275} (\bibinfo {year}
  {2015})}\BibitemShut {NoStop}%
\bibitem [{\citenamefont {Lostaglio}\ \emph
  {et~al.}(2015{\natexlab{a}})\citenamefont {Lostaglio}, \citenamefont
  {Jennings},\ and\ \citenamefont {Rudolph}}]{Rudolph214}%
  \BibitemOpen
  \bibfield  {author} {\bibinfo {author} {\bibfnamefont {M.}~\bibnamefont
  {Lostaglio}}, \bibinfo {author} {\bibfnamefont {D.}~\bibnamefont {Jennings}},
  \ and\ \bibinfo {author} {\bibfnamefont {T.}~\bibnamefont {Rudolph}},\ }\href
  {http://dx.doi.org/10.1038/ncomms7383} {\bibfield  {journal} {\bibinfo
  {journal} {Nature Communications}\ }\textbf {\bibinfo {volume} {6}} (\bibinfo
  {year} {2015}{\natexlab{a}})}\BibitemShut {NoStop}%
\bibitem [{\citenamefont {Lostaglio}\ \emph
  {et~al.}(2015{\natexlab{b}})\citenamefont {Lostaglio}, \citenamefont
  {Korzekwa}, \citenamefont {Jennings},\ and\ \citenamefont
  {Rudolph}}]{Rudolph114}%
  \BibitemOpen
  \bibfield  {author} {\bibinfo {author} {\bibfnamefont {M.}~\bibnamefont
  {Lostaglio}}, \bibinfo {author} {\bibfnamefont {K.}~\bibnamefont {Korzekwa}},
  \bibinfo {author} {\bibfnamefont {D.}~\bibnamefont {Jennings}}, \ and\
  \bibinfo {author} {\bibfnamefont {T.}~\bibnamefont {Rudolph}},\ }\href
  {\doibase 10.1103/PhysRevX.5.021001} {\bibfield  {journal} {\bibinfo
  {journal} {Phys. Rev. X}\ }\textbf {\bibinfo {volume} {5}},\ \bibinfo {pages}
  {021001} (\bibinfo {year} {2015}{\natexlab{b}})}\BibitemShut {NoStop}%
\bibitem [{\citenamefont {Faist}\ \emph {et~al.}(2015)\citenamefont {Faist},
  \citenamefont {Oppenheim},\ and\ \citenamefont {Renner}}]{Faist2015}%
  \BibitemOpen
  \bibfield  {author} {\bibinfo {author} {\bibfnamefont {P.}~\bibnamefont
  {Faist}}, \bibinfo {author} {\bibfnamefont {J.}~\bibnamefont {Oppenheim}}, \
  and\ \bibinfo {author} {\bibfnamefont {R.}~\bibnamefont {Renner}},\ }\href
  {http://stacks.iop.org/1367-2630/17/i=4/a=043003} {\bibfield  {journal}
  {\bibinfo  {journal} {New Journal of Physics}\ }\textbf {\bibinfo {volume}
  {17}},\ \bibinfo {pages} {043003} (\bibinfo {year} {2015})}\BibitemShut
  {NoStop}%
\bibitem [{\citenamefont {Mandelbrot}(1982)}]{Mandelbrot1982}%
  \BibitemOpen
  \bibfield  {author} {\bibinfo {author} {\bibfnamefont {B.~B.}\ \bibnamefont
  {Mandelbrot}},\ }\href@noop {} {\emph {\bibinfo {title} {The Fractal Geometry
  of Nature}}}\ (\bibinfo  {publisher} {W. H. Freeman and company New York},\
  \bibinfo {year} {1982})\BibitemShut {NoStop}%
\bibitem [{\citenamefont {Mandelbrot}(1999)}]{Mandelbrot1999}%
  \BibitemOpen
  \bibfield  {author} {\bibinfo {author} {\bibfnamefont {B.~B.}\ \bibnamefont
  {Mandelbrot}},\ }\href {\doibase 10.1007/978-1-4612-2150-0} {\emph {\bibinfo
  {title} {Multifractals and 1/f Noise}}}\ (\bibinfo  {publisher}
  {Springer-Verlag New York},\ \bibinfo {year} {1999})\BibitemShut {NoStop}%
\bibitem [{\citenamefont {Jizba}\ and\ \citenamefont
  {Arimitsu}(2004)}]{Jizba2004}%
  \BibitemOpen
  \bibfield  {author} {\bibinfo {author} {\bibfnamefont {P.}~\bibnamefont
  {Jizba}}\ and\ \bibinfo {author} {\bibfnamefont {T.}~\bibnamefont
  {Arimitsu}},\ }\href {\doibase http://dx.doi.org/10.1016/j.aop.2004.01.002}
  {\bibfield  {journal} {\bibinfo  {journal} {Annals of Physics}\ }\textbf
  {\bibinfo {volume} {312}},\ \bibinfo {pages} {17 } (\bibinfo {year}
  {2004})}\BibitemShut {NoStop}%
\bibitem [{\citenamefont {{Tomamichel}}()}]{Tomamichel2015}%
  \BibitemOpen
  \bibfield  {author} {\bibinfo {author} {\bibfnamefont {M.}~\bibnamefont
  {{Tomamichel}}},\ }\href@noop {} {}\Eprint {http://arxiv.org/abs/1504.00233}
  {arXiv:1504.00233} \BibitemShut {NoStop}%
\bibitem [{\citenamefont {R\'{e}nyi}(1961)}]{Renyi1961}%
  \BibitemOpen
  \bibfield  {author} {\bibinfo {author} {\bibfnamefont {A.}~\bibnamefont
  {R\'{e}nyi}},\ }\enquote {\bibinfo {title} {On measures of entropy and
  information},}\ in\ \href {http://projecteuclid.org/euclid.bsmsp/1200512181}
  {\emph {\bibinfo {booktitle} {Proceedings of the Fourth Berkeley Symposium on
  Mathematical Statistics and Probability}}}\ (\bibinfo  {publisher}
  {University of California Press},\ \bibinfo {year} {1961})\ pp.\ \bibinfo
  {pages} {547--561}\BibitemShut {NoStop}%
\bibitem [{\citenamefont {R\'{e}nyi}(1965)}]{Renyi1965}%
  \BibitemOpen
  \bibfield  {author} {\bibinfo {author} {\bibfnamefont {A.}~\bibnamefont
  {R\'{e}nyi}},\ }\href {http://www.jstor.org/stable/1401301} {\bibfield
  {journal} {\bibinfo  {journal} {Review of the International Statistical
  Institute}\ }\textbf {\bibinfo {volume} {33}},\ \bibinfo {pages} {1}
  (\bibinfo {year} {1965})}\BibitemShut {NoStop}%
\bibitem [{\citenamefont {Lenzi}\ \emph
  {et~al.}(2000{\natexlab{a}})\citenamefont {Lenzi}, \citenamefont {Mendes},\
  and\ \citenamefont {da~Silva}}]{Lenzi2000}%
  \BibitemOpen
  \bibfield  {author} {\bibinfo {author} {\bibfnamefont {E.}~\bibnamefont
  {Lenzi}}, \bibinfo {author} {\bibfnamefont {R.}~\bibnamefont {Mendes}}, \
  and\ \bibinfo {author} {\bibfnamefont {L.}~\bibnamefont {da~Silva}},\ }\href
  {\doibase http://dx.doi.org/10.1016/S0378-4371(00)00007-8} {\bibfield
  {journal} {\bibinfo  {journal} {Physica A: Statistical Mechanics and its
  Applications}\ }\textbf {\bibinfo {volume} {280}},\ \bibinfo {pages} {337 }
  (\bibinfo {year} {2000}{\natexlab{a}})}\BibitemShut {NoStop}%
\bibitem [{\citenamefont {Horodecki}\ \emph {et~al.}(1996)\citenamefont
  {Horodecki}, \citenamefont {Horodecki},\ and\ \citenamefont
  {Horodecki}}]{Horodeckia1996}%
  \BibitemOpen
  \bibfield  {author} {\bibinfo {author} {\bibfnamefont {R.}~\bibnamefont
  {Horodecki}}, \bibinfo {author} {\bibfnamefont {P.}~\bibnamefont
  {Horodecki}}, \ and\ \bibinfo {author} {\bibfnamefont {M.}~\bibnamefont
  {Horodecki}},\ }\href {\doibase
  http://dx.doi.org/10.1016/0375-9601(95)00930-2} {\bibfield  {journal}
  {\bibinfo  {journal} {Physics Letters A}\ }\textbf {\bibinfo {volume}
  {210}},\ \bibinfo {pages} {377 } (\bibinfo {year} {1996})}\BibitemShut
  {NoStop}%
\bibitem [{\citenamefont {Plastino}\ and\ \citenamefont
  {Plastino}(1997)}]{Plastino1997}%
  \BibitemOpen
  \bibfield  {author} {\bibinfo {author} {\bibfnamefont {A.}~\bibnamefont
  {Plastino}}\ and\ \bibinfo {author} {\bibfnamefont {A.}~\bibnamefont
  {Plastino}},\ }\href {\doibase
  http://dx.doi.org/10.1016/S0375-9601(96)00942-5} {\bibfield  {journal}
  {\bibinfo  {journal} {Physics Letters A}\ }\textbf {\bibinfo {volume}
  {226}},\ \bibinfo {pages} {257 } (\bibinfo {year} {1997})}\BibitemShut
  {NoStop}%
\bibitem [{\citenamefont {Rajagopal}\ \emph {et~al.}(2002)\citenamefont
  {Rajagopal}, \citenamefont {Rendell},\ and\ \citenamefont
  {Abe}}]{Rajagopal2002}%
  \BibitemOpen
  \bibfield  {author} {\bibinfo {author} {\bibfnamefont {A.~K.}\ \bibnamefont
  {Rajagopal}}, \bibinfo {author} {\bibfnamefont {R.~W.}\ \bibnamefont
  {Rendell}}, \ and\ \bibinfo {author} {\bibfnamefont {S.}~\bibnamefont
  {Abe}},\ }\href {\doibase http://dx.doi.org/10.1063/1.1523780} {\bibfield
  {journal} {\bibinfo  {journal} {AIP Conference Proceedings}\ }\textbf
  {\bibinfo {volume} {643}},\ \bibinfo {pages} {47} (\bibinfo {year}
  {2002})}\BibitemShut {NoStop}%
\bibitem [{\citenamefont {Oono}\ and\ \citenamefont
  {Paniconi}(1998)}]{Oono1998}%
  \BibitemOpen
  \bibfield  {author} {\bibinfo {author} {\bibfnamefont {Y.}~\bibnamefont
  {Oono}}\ and\ \bibinfo {author} {\bibfnamefont {M.}~\bibnamefont
  {Paniconi}},\ }\href {\doibase 10.1143/PTPS.130.29} {\bibfield  {journal}
  {\bibinfo  {journal} {Progress of Theoretical Physics Supplement}\ }\textbf
  {\bibinfo {volume} {130}},\ \bibinfo {pages} {29} (\bibinfo {year}
  {1998})}\BibitemShut {NoStop}%
\bibitem [{\citenamefont {Hatano}\ and\ \citenamefont {Sasa}(2001)}]{Sasa2001}%
  \BibitemOpen
  \bibfield  {author} {\bibinfo {author} {\bibfnamefont {T.}~\bibnamefont
  {Hatano}}\ and\ \bibinfo {author} {\bibfnamefont {S.}~\bibnamefont {Sasa}},\
  }\href {\doibase 10.1103/PhysRevLett.86.3463} {\bibfield  {journal} {\bibinfo
   {journal} {Phys. Rev. Lett.}\ }\textbf {\bibinfo {volume} {86}},\ \bibinfo
  {pages} {3463} (\bibinfo {year} {2001})}\BibitemShut {NoStop}%
\bibitem [{\citenamefont {Seifert}(2012)}]{Seifert2012}%
  \BibitemOpen
  \bibfield  {author} {\bibinfo {author} {\bibfnamefont {U.}~\bibnamefont
  {Seifert}},\ }\href {http://stacks.iop.org/0034-4885/75/i=12/a=126001}
  {\bibfield  {journal} {\bibinfo  {journal} {Reports on Progress in Physics}\
  }\textbf {\bibinfo {volume} {75}},\ \bibinfo {pages} {126001} (\bibinfo
  {year} {2012})}\BibitemShut {NoStop}%
\bibitem [{\citenamefont {{Deffner}}\ and\ \citenamefont
  {{Lutz}}()}]{Deffner2012}%
  \BibitemOpen
  \bibfield  {author} {\bibinfo {author} {\bibfnamefont {S.}~\bibnamefont
  {{Deffner}}}\ and\ \bibinfo {author} {\bibfnamefont {E.}~\bibnamefont
  {{Lutz}}},\ }\href@noop {} {}\Eprint {http://arxiv.org/abs/1201.3888}
  {arXiv:1201.3888} \BibitemShut {NoStop}%
\bibitem [{\citenamefont {Lennert}\ \emph {et~al.}(2013)\citenamefont
  {Lennert}, \citenamefont {Dupuis}, \citenamefont {Szehr}, \citenamefont
  {Fehr},\ and\ \citenamefont {Tomamichel}}]{Tomamichel2013}%
  \BibitemOpen
  \bibfield  {author} {\bibinfo {author} {\bibfnamefont {M.}~\bibnamefont
  {Lennert}}, \bibinfo {author} {\bibfnamefont {F.}~\bibnamefont {Dupuis}},
  \bibinfo {author} {\bibfnamefont {O.}~\bibnamefont {Szehr}}, \bibinfo
  {author} {\bibfnamefont {S.}~\bibnamefont {Fehr}}, \ and\ \bibinfo {author}
  {\bibfnamefont {M.}~\bibnamefont {Tomamichel}},\ }\href {\doibase
  http://dx.doi.org/10.1063/1.4838856} {\bibfield  {journal} {\bibinfo
  {journal} {Journal of Mathematical Physics}\ }\textbf {\bibinfo {volume}
  {54}},\ \bibinfo {eid} {122203} (\bibinfo {year} {2013})}\BibitemShut
  {NoStop}%
\bibitem [{\citenamefont {Wilde}\ \emph {et~al.}(2014)\citenamefont {Wilde},
  \citenamefont {Winter},\ and\ \citenamefont {Yang}}]{Wilde2014}%
  \BibitemOpen
  \bibfield  {author} {\bibinfo {author} {\bibfnamefont {M.}~\bibnamefont
  {Wilde}}, \bibinfo {author} {\bibfnamefont {A.}~\bibnamefont {Winter}}, \
  and\ \bibinfo {author} {\bibfnamefont {D.}~\bibnamefont {Yang}},\ }\href
  {\doibase 10.1007/s00220-014-2122-x} {\bibfield  {journal} {\bibinfo
  {journal} {Communications in Mathematical Physics}\ }\textbf {\bibinfo
  {volume} {331}},\ \bibinfo {pages} {593} (\bibinfo {year}
  {2014})}\BibitemShut {NoStop}%
\bibitem [{\citenamefont {Rajagopal}\ \emph {et~al.}(2014)\citenamefont
  {Rajagopal}, \citenamefont {Sudha}, \citenamefont {Nayak},\ and\
  \citenamefont {Devi}}]{Rajagopal2014}%
  \BibitemOpen
  \bibfield  {author} {\bibinfo {author} {\bibfnamefont {A.~K.}\ \bibnamefont
  {Rajagopal}}, \bibinfo {author} {\bibnamefont {Sudha}}, \bibinfo {author}
  {\bibfnamefont {A.~S.}\ \bibnamefont {Nayak}}, \ and\ \bibinfo {author}
  {\bibfnamefont {A.~R.~U.}\ \bibnamefont {Devi}},\ }\href {\doibase
  10.1103/PhysRevA.89.012331} {\bibfield  {journal} {\bibinfo  {journal} {Phys.
  Rev. A}\ }\textbf {\bibinfo {volume} {89}},\ \bibinfo {pages} {012331}
  (\bibinfo {year} {2014})}\BibitemShut {NoStop}%
\bibitem [{\citenamefont {{Nayak}}\ \emph {et~al.}()\citenamefont {{Nayak}},
  \citenamefont {{Sudha}}, \citenamefont {{Rajagopal}},\ and\ \citenamefont
  {{Usha Devi}}}]{Nayak2014}%
  \BibitemOpen
  \bibfield  {author} {\bibinfo {author} {\bibfnamefont {A.~S.}\ \bibnamefont
  {{Nayak}}}, \bibinfo {author} {\bibnamefont {{Sudha}}}, \bibinfo {author}
  {\bibfnamefont {A.~K.}\ \bibnamefont {{Rajagopal}}}, \ and\ \bibinfo {author}
  {\bibfnamefont {A.~R.}\ \bibnamefont {{Usha Devi}}},\ }\href@noop {}
  {}\bibinfo {note} {To appear in Physica A: Statistical Mechanics and its
  Applications (2015)},\ \Eprint {http://arxiv.org/abs/1407.4386}
  {arXiv:1407.4386} \BibitemShut {NoStop}%
\bibitem [{\citenamefont {Misra}\ \emph {et~al.}(2015)\citenamefont {Misra},
  \citenamefont {Biswas}, \citenamefont {Pati}, \citenamefont {Sen(De)},\ and\
  \citenamefont {Sen}}]{Avijit2014}%
  \BibitemOpen
  \bibfield  {author} {\bibinfo {author} {\bibfnamefont {A.}~\bibnamefont
  {Misra}}, \bibinfo {author} {\bibfnamefont {A.}~\bibnamefont {Biswas}},
  \bibinfo {author} {\bibfnamefont {A.~K.}\ \bibnamefont {Pati}}, \bibinfo
  {author} {\bibfnamefont {A.}~\bibnamefont {Sen(De)}}, \ and\ \bibinfo
  {author} {\bibfnamefont {U.}~\bibnamefont {Sen}},\ }\href {\doibase
  10.1103/PhysRevE.91.052125} {\bibfield  {journal} {\bibinfo  {journal} {Phys.
  Rev. E}\ }\textbf {\bibinfo {volume} {91}},\ \bibinfo {pages} {052125}
  (\bibinfo {year} {2015})}\BibitemShut {NoStop}%
\bibitem [{\citenamefont {Tomamichel}()}]{Tomamichel2012}%
  \BibitemOpen
  \bibfield  {author} {\bibinfo {author} {\bibfnamefont {M.}~\bibnamefont
  {Tomamichel}},\ }\href@noop {} {}\Eprint {http://arxiv.org/abs/1203.2142}
  {arXiv:1203.2142} \BibitemShut {NoStop}%
\bibitem [{\citenamefont {{Renner}}(2005)}]{RennerPhD2005}%
  \BibitemOpen
  \bibfield  {author} {\bibinfo {author} {\bibfnamefont {R.}~\bibnamefont
  {{Renner}}},\ }\emph {\bibinfo {title} {{Security of Quantum Key
  Distribution}}},\ \href {http://arxiv.org/abs/quant-ph/0512258} {Ph.D.
  thesis},\ \bibinfo  {school} {ETH Zurich} (\bibinfo {year}
  {2005})\BibitemShut {NoStop}%
\bibitem [{\citenamefont {Datta}(2009)}]{Nilanjana2009}%
  \BibitemOpen
  \bibfield  {author} {\bibinfo {author} {\bibfnamefont {N.}~\bibnamefont
  {Datta}},\ }\href {\doibase 10.1109/TIT.2009.2018325} {\bibfield  {journal}
  {\bibinfo  {journal} {IEEE Transactions on Information Theory}\ }\textbf
  {\bibinfo {volume} {55}},\ \bibinfo {pages} {2816} (\bibinfo {year}
  {2009})}\BibitemShut {NoStop}%
\bibitem [{\citenamefont {Callen}(1985)}]{Callen1985}%
  \BibitemOpen
  \bibfield  {author} {\bibinfo {author} {\bibfnamefont {H.}~\bibnamefont
  {Callen}},\ }\href@noop {} {\emph {\bibinfo {title} {Thermodynamics and an
  introduction to thermostatistics}}}\ (\bibinfo  {publisher} {Wiley},\
  \bibinfo {address} {New York},\ \bibinfo {year} {1985})\BibitemShut {NoStop}%
\bibitem [{\citenamefont {Gemmer}\ \emph {et~al.}(2004)\citenamefont {Gemmer},
  \citenamefont {Michel},\ and\ \citenamefont {Mahler}}]{Gemmer2004}%
  \BibitemOpen
  \bibfield  {author} {\bibinfo {author} {\bibfnamefont {J.}~\bibnamefont
  {Gemmer}}, \bibinfo {author} {\bibfnamefont {M.}~\bibnamefont {Michel}}, \
  and\ \bibinfo {author} {\bibfnamefont {G.}~\bibnamefont {Mahler}},\ }\href
  {\doibase 10.1007/b98082} {\emph {\bibinfo {title} {Quantum
  Thermodynamics}}}\ (\bibinfo  {publisher} {Springer Berlin Heidelberg},\
  \bibinfo {year} {2004})\BibitemShut {NoStop}%
\bibitem [{\citenamefont {Petz}(1986)}]{Petz1986}%
  \BibitemOpen
  \bibfield  {author} {\bibinfo {author} {\bibfnamefont {D.}~\bibnamefont
  {Petz}},\ }\href {\doibase http://dx.doi.org/10.1016/0034-4877(86)90067-4}
  {\bibfield  {journal} {\bibinfo  {journal} {Reports on Mathematical Physics}\
  }\textbf {\bibinfo {volume} {23}},\ \bibinfo {pages} {57 } (\bibinfo {year}
  {1986})}\BibitemShut {NoStop}%
\bibitem [{\citenamefont {Abe}\ and\ \citenamefont
  {Rajagopal}(2003)}]{Abe2003}%
  \BibitemOpen
  \bibfield  {author} {\bibinfo {author} {\bibfnamefont {S.}~\bibnamefont
  {Abe}}\ and\ \bibinfo {author} {\bibfnamefont {A.~K.}\ \bibnamefont
  {Rajagopal}},\ }\href {\doibase 10.1103/PhysRevLett.91.120601} {\bibfield
  {journal} {\bibinfo  {journal} {Phys. Rev. Lett.}\ }\textbf {\bibinfo
  {volume} {91}},\ \bibinfo {pages} {120601} (\bibinfo {year}
  {2003})}\BibitemShut {NoStop}%
\bibitem [{\citenamefont {Beigi}(2013)}]{Beigi2013}%
  \BibitemOpen
  \bibfield  {author} {\bibinfo {author} {\bibfnamefont {S.}~\bibnamefont
  {Beigi}},\ }\href {\doibase http://dx.doi.org/10.1063/1.4838855} {\bibfield
  {journal} {\bibinfo  {journal} {Journal of Mathematical Physics}\ }\textbf
  {\bibinfo {volume} {54}},\ \bibinfo {eid} {122202} (\bibinfo {year}
  {2013})}\BibitemShut {NoStop}%
\bibitem [{\citenamefont {Frank}\ and\ \citenamefont {Lieb}(2013)}]{Frank2013}%
  \BibitemOpen
  \bibfield  {author} {\bibinfo {author} {\bibfnamefont {R.~L.}\ \bibnamefont
  {Frank}}\ and\ \bibinfo {author} {\bibfnamefont {E.~H.}\ \bibnamefont
  {Lieb}},\ }\href {\doibase http://dx.doi.org/10.1063/1.4838835} {\bibfield
  {journal} {\bibinfo  {journal} {Journal of Mathematical Physics}\ }\textbf
  {\bibinfo {volume} {54}},\ \bibinfo {eid} {122201} (\bibinfo {year}
  {2013})}\BibitemShut {NoStop}%
\bibitem [{\citenamefont {Rajagopal}\ \emph {et~al.}(1998)\citenamefont
  {Rajagopal}, \citenamefont {Mendes},\ and\ \citenamefont
  {Lenzi}}]{Rajagopal1998}%
  \BibitemOpen
  \bibfield  {author} {\bibinfo {author} {\bibfnamefont {A.~K.}\ \bibnamefont
  {Rajagopal}}, \bibinfo {author} {\bibfnamefont {R.~S.}\ \bibnamefont
  {Mendes}}, \ and\ \bibinfo {author} {\bibfnamefont {E.~K.}\ \bibnamefont
  {Lenzi}},\ }\href {\doibase 10.1103/PhysRevLett.80.3907} {\bibfield
  {journal} {\bibinfo  {journal} {Phys. Rev. Lett.}\ }\textbf {\bibinfo
  {volume} {80}},\ \bibinfo {pages} {3907} (\bibinfo {year}
  {1998})}\BibitemShut {NoStop}%
\bibitem [{\citenamefont {Lenzi}\ \emph
  {et~al.}(2000{\natexlab{b}})\citenamefont {Lenzi}, \citenamefont {Mendes},\
  and\ \citenamefont {Rajagopal}}]{LenziB2000}%
  \BibitemOpen
  \bibfield  {author} {\bibinfo {author} {\bibfnamefont {E.~K.}\ \bibnamefont
  {Lenzi}}, \bibinfo {author} {\bibfnamefont {R.~S.}\ \bibnamefont {Mendes}}, \
  and\ \bibinfo {author} {\bibfnamefont {A.~K.}\ \bibnamefont {Rajagopal}},\
  }\href {\doibase http://dx.doi.org/10.1016/S0378-4371(00)00364-2} {\bibfield
  {journal} {\bibinfo  {journal} {Physica A: Statistical Mechanics and its
  Applications}\ }\textbf {\bibinfo {volume} {286}},\ \bibinfo {pages} {503 }
  (\bibinfo {year} {2000}{\natexlab{b}})}\BibitemShut {NoStop}%
\bibitem [{\citenamefont {Martin}\ and\ \citenamefont
  {Schwinger}(1959)}]{Martin1959}%
  \BibitemOpen
  \bibfield  {author} {\bibinfo {author} {\bibfnamefont {P.~C.}\ \bibnamefont
  {Martin}}\ and\ \bibinfo {author} {\bibfnamefont {J.}~\bibnamefont
  {Schwinger}},\ }\href {\doibase 10.1103/PhysRev.115.1342} {\bibfield
  {journal} {\bibinfo  {journal} {Phys. Rev.}\ }\textbf {\bibinfo {volume}
  {115}},\ \bibinfo {pages} {1342} (\bibinfo {year} {1959})}\BibitemShut
  {NoStop}%
\bibitem [{\citenamefont {Brand\~ao}\ \emph {et~al.}(2013)\citenamefont
  {Brand\~ao}, \citenamefont {Horodecki}, \citenamefont {Oppenheim},
  \citenamefont {Renes},\ and\ \citenamefont {Spekkens}}]{Fernandoa2013}%
  \BibitemOpen
  \bibfield  {author} {\bibinfo {author} {\bibfnamefont {F.~G. S.~L.}\
  \bibnamefont {Brand\~ao}}, \bibinfo {author} {\bibfnamefont {M.}~\bibnamefont
  {Horodecki}}, \bibinfo {author} {\bibfnamefont {J.}~\bibnamefont
  {Oppenheim}}, \bibinfo {author} {\bibfnamefont {J.~M.}\ \bibnamefont
  {Renes}}, \ and\ \bibinfo {author} {\bibfnamefont {R.~W.}\ \bibnamefont
  {Spekkens}},\ }\href {\doibase 10.1103/PhysRevLett.111.250404} {\bibfield
  {journal} {\bibinfo  {journal} {Phys. Rev. Lett.}\ }\textbf {\bibinfo
  {volume} {111}},\ \bibinfo {pages} {250404} (\bibinfo {year}
  {2013})}\BibitemShut {NoStop}%
\end{thebibliography}%

\end{document}